# Universal Quantized Berry-Dipole Flat Bands

Qingyang Mo[1,2] and Shuang Zhang[1,2,3,4,5 †]

[1] *New Cornerstone Science Laboratory, Department of Physics, The University of Hong Kong; 999077, Hong Kong, China*
[2] *State Key Laboratory of Optical Quantum Materials, The University of Hong Kong, 999077 Hong Kong, China*
[3] *Department of Electrical and Electronic Engineering, The University of Hong Kong, 999077 Hong Kong, China*
[4] *Quantum Science Center of Guangdong-Hong Kong-Macao Great Bay Area, 3 Binlang Road, Shenzhen, China*
[5] *Materials Innovation Institute for Life Sciences and Energy (MILES), HKU-SIRI, Shenzhen, People's Republic of China*

Perfectly flat bands with nontrivial quantum geometry have emerged as a frontier for exotic topological phenomena and superconductors. Here, we unveil a universal family of quantized Berry-dipole flat bands in chiral-symmetric $(2n+1)$-band systems, where the central perfectly flat band carries a Berry-dipole moment $d = n$, with $n$ an arbitrary integer, while preserving zero Chern number. We construct explicit lattice models to showcase three topological phenomena characterized by the Berry-dipole moment: a flat-band returning pump featuring bidirectional, soliton-like displacement of Wannier centers by exactly $n$ unit cells per half cycle, a "dipolar" Haldane phase diagram arising from the competition between time-reversal and parity symmetries, and $n$ pairs of bulk helical zero modes whose existence depends on the orientation of pseudomagnetic field. Our findings establish a universal framework for the topology beyond Chern class in perfectly flat bands and provide a tunable platform for exploring quantum geometry and interaction-driven phases.

The search for topological quantum matter has been revolutionized by topological flat bands, where nontrivial topology coexists with quenched kinetic energy, providing an ideal setting for strong correlations and exotic many-body phases. Paradigmatic examples include the flat Landau levels of quantum Hall systems, which host the fractional quantum Hall effect (FQHE) [1–3], and the moiré flat bands in twisted graphene heterostructures, which have recently revealed the fractional quantum anomalous Hall effect (FQAHE) at zero magnetic field [4–11]. These platforms demonstrate how flat dispersion and topological invariants can cooperate to stabilize fractionalized excitations and interaction-driven topological order.

A conceptually distinct class is that of perfectly flat bands with strictly zero bandwidth, enforced by geometric frustration or chiral symmetry. Historically viewed as topologically trivial at the single-particle level, interest in such bands has centered on their extensive degeneracy and localized wave functions, rather than on any topological character [12–16].

This conventional picture has been transformed by the broader perspective of quantum geometry. Beyond the Chern number (Berry monopole), higher multipole moments of the Berry curvature, such as dipoles and quadrupoles, can encode delicate or fragile topology even when the Chern number vanishes [17–22]. Furthermore, the quantum metric in flat bands, which measures the distance between Bloch states in momentum space, can induce anomalous dispersion [23–25], modified superconducting coherence lengths [26,27], and has been proposed to help stabilize fractionalized phases [28,29]. These developments raise a clear question: *can a perfectly flat band with zero Chern number be characterized by a universal, quantized invariant that grows to arbitrary integers and unifies multiple topological responses?*

Here, we answer this question by introducing a universal family of quantized Berry-dipole flat bands. In our $(2n+1)$-band models with chiral symmetry, the central perfectly flat band carries an integer Berry-dipole moment, $n = 1, 2, 3, ...$, while retaining zero Chern number. This quantized vector invariant governs several universal phenomena that transcend dimensions: (i) a one-dimensional (1D) generalized returning Thouless pump (RTP) with Wannier centers moving bidirectionally by exactly $n$ unit cells; (ii) a 2D "dipolar" Haldane insulator that hosts $n$ pairs of helical edge states, arising from competition between time-reversal ($T$) and parity ($P$) symmetries; and (iii) $n$ pairs of bulk helical zero modes whose existence is controlled by the orientation of a pseudomagnetic field. Our work establishes the quantized Berry dipole as a fundamental topological index for perfect flat bands, opening a tunable platform to explore delicate topology, quantum geometry, and strong correlations.

*Continuum model*. We begin by constructing a continuum model that hosts a perfectly flat band with a quantized Berry-dipole moment. The Hamiltonian acts on a basis of $(2n + 1)$ bands ($n = 1,2,3, ...$) and takes a chiral-

symmetric, tridiagonal form:

$$H_{2n+1} = \begin{pmatrix} 0 & d_1 & 0 & 0 & 0 & \cdots & 0 \\ d_1 & 0 & d_2 & 0 & 0 & \cdots & 0 \\ 0 & d_2^* & 0 & d_1 & 0 & \cdots & 0 \\ 0 & 0 & d_1 & 0 & d_2 & \cdots & 0 \\ 0 & 0 & 0 & d_2^* & 0 & \cdots & 0 \\ \vdots & \vdots & \vdots & \vdots & \vdots & \ddots & \vdots \\ 0 & 0 & 0 & 0 & 0 & \cdots & 0 \end{pmatrix}, \quad (1)$$

where $d_1 = k_z, d_2 = k_x - ik_y$. These Hamiltonian respects chiral symmetry $\Gamma H \Gamma^\dagger = -H$, with $\Gamma = \text{diag}\{1, -1, 1, -1, \ldots\}$), which guarantees a symmetric spectrum and enforces a perfectly flat band at zero energy. As shown in Fig. 1(a) for $n = 1,2,3$, the band spectrum features a $(2n+1)$-fold degeneracy at $k = 0$ and a strictly dispersionless band (highlighted in red). The flat band carries a quantized Berry-dipole moment $\vec{d} = (0,0,n)$ oriented along $k_z$, a direct consequence of the Hamiltonian's rotational symmetry. The integer $n$ is the topological invariant computed via the flux of Berry curvature $\Omega(k)$ through the upper half $S^+$ ($k_z > 0$) of the sphere enclosing the degeneracy:

$$\frac{1}{2\pi} \int_{S^+} \Omega \cdot dS = n, \quad n \in \{1, 2, 3, \ldots\}. \quad (2)$$

In Fig. 1(b), we visualize the Berry curvature distribution on a sphere surrounding the degeneracy for $n = 1,2,3$, clearly revealing a dipole pattern whose strength scales with the integer $n$.

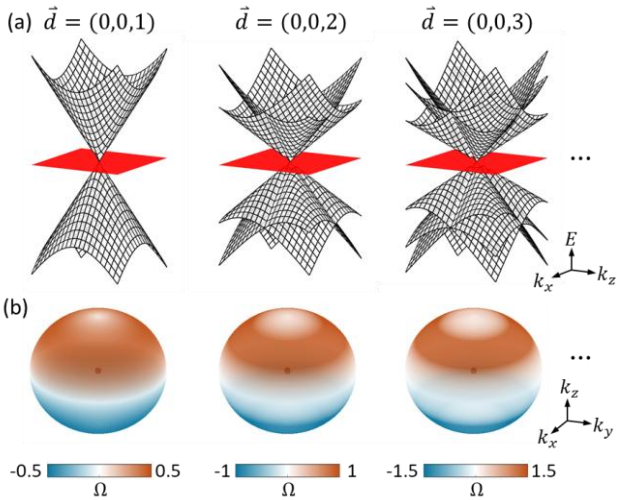

FIG. 1. Band structures and Berry curvature distributions for quantized Berry-dipole flat bands in $(2n+1)$-band systems. (a) Energy bands with a perfectly flat central band (red) and $(2n+1)$-fold degeneracy at $k = 0$. (b) Berry curvature on a sphere enclosing the degeneracy, showing dipole patterns scaling with integer $n$ ($n = 1,2,3$ examples).

*Generalized RTP.* We now demonstrate that the quantized Berry-dipole moment governs a generalized RTP [30,31]. In contrast to conventional Thouless pumps that rely on dispersive bands, our scheme presents the first realization of a quantized topological pump within a perfectly flat band, where the band remains strictly dispersionless throughout the entire pumping cycle. To construct a minimal lattice realization, we consider a family of Rice-Mele-type chains with $(2n+1)$ sites (labeled $1, 2, \ldots, 2n+1$) per unit cell, as shown in Figs. 2(a-c) for $n = 1,2,3$. The Hamiltonian reads:

$$\hat{H}(t) = \sum_{i \in [1,n], j} \{v(t) \hat{a}^\dagger_{2i-1,j} \hat{a}_{2i,j} + w_1(t) \hat{a}^\dagger_{2i,j} \hat{a}_{2i+1,j}$$
$$+ w_2(t) \hat{a}^\dagger_{2i,j} \hat{a}_{2i+1,j+1} + + h.c.\}, \quad (3)$$

where $\hat{a}^\dagger_{i,j} (\hat{a}_{i,j})$ is the creation (annihilation) operator acting on the $i$-th site of the $j$-th unit cell. The time-dependent couplings are modulated as $v = v_0 \sin(2\pi t/T)$ and $w_{1,2} = w_0 \pm \Delta \cos(2\pi t/T)$, where T is the pumping period. After Fourier transformation, the dynamic 1D k-space Hamiltonian satisfies Eq. (1) with

$$d_1 = v_0 \sin\left(\frac{2\pi t}{T}\right),$$

$$d_2 = 2w_0 \cos\left(\frac{k}{2}\right) - 2i\Delta \sin\left(\frac{k}{2}\right) \cos\left(\frac{2\pi t}{T}\right). \quad (4)$$

As the pumping parameter $t$ evolves a full period from 0 to $T$, the trajectory in parameter space winds once around the Berry-dipole degeneracy point, guaranteeing that the central flat band carries the quantized dipole moment $n$.

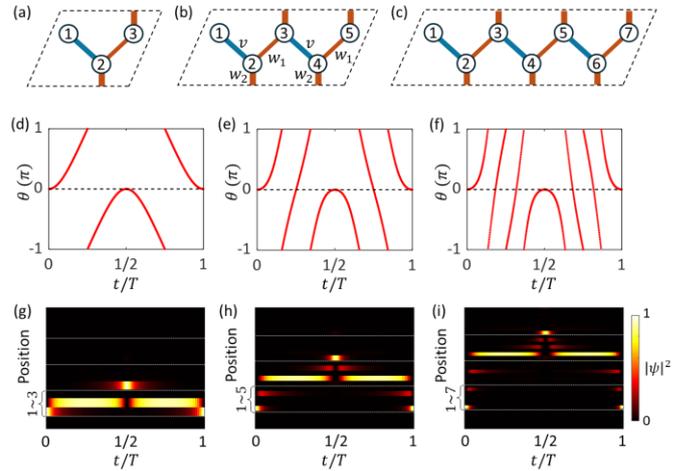

FIG. 2. Generalized returning Thouless pump in a Rice-Mele-type lattice model. (a-c) The scheme of lattice model for $n = 1,2,3$. Blue bonds denote the coupling $v = v_0 \sin(2\pi t/T)$, while thin (thick) red bonds denote the intra-(inter-)coupling $w_{1,2} = w_0 \pm \Delta \cos(2\pi t/T)$, respectively. (d-f) Berry phase as a function of pumping parameter $t/T$ for $n = 1,2,3$, where $\Delta = 0.5 w_0$ and $v_0 = w_0$. (g-i) Time evolution of the probability density $|\psi(t)|^2$ starting from a localized state on site A for $n = 1,2,3$, where $\Delta = w_0$ and $v_0 = 5 w_0$.

The dipole topology manifests in the cumulated Berry phase of the flat band. We compute the Berry phase $\theta(t)$ (equivalent to the Wannier center modulo $2\pi$) via Wilson loop method along the band during the cycle, shown in Figs. 2(d-f) [see more details about the Wilson loop method in Supplementary materials (SM), section 1]. Remarkably, $\theta(t)$ accumulates a phase of $2n\pi$ during the first half-cycle and $-2n\pi$ during the second, corresponding to the Wannier center first advancing forward exactly $n$ unit cells and then returning to its initial position over a full cycle. This behavior is a hallmark of a quantized returning pump whose winding number is set by the dipole order $n$. Real-time dynamics confirm this picture. Starting from a state localized on site 1, we evolve the system under the slowly modulated Hamiltonian. The time-dependent probability density, plotted in Figs. 2(g–i), exhibits a clear bidirectional soliton-like propagation: for each $n$, the wave packet advances by $n$ unit cells in the first half-cycle and retraces its path in the second half, returning exactly to its initial location. Crucially, the strictly flatness of the band ensures a sharply localized, "soliton-like" propagating profile in the linear regime, which stands in sharp contrast to the diffusive broadening typical of pumps in dispersive bands.

*Dipolar Haldane model*. We now construct a static two-dimensional insulator, called "dipolar Haldane model", to demonstrate the bulk-boundary correspondence dictated by the quantized Berry dipole. Fig. 3(a) shows the $n$-layer-stacked 2D lattice model, with each unit cell comprising $(2n+1)$ sites (yellow spheres), where $n$ defines the quantized Berry-dipole moment. The model contains three types of couplings: diagonal couplings $t_0$ (dashed bonds) preserving both $T$ and $P$ symmetries, complex diagonal couplings $t_1 e^{i\phi}$ (arrowed bonds) that break $T$ symmetry, and leg couplings $t_0 \pm \Delta$ (red/blue bonds) that break $P$ symmetry. The 2D k-space Hamiltonian satisfies Eq. (1) with

$$d_1 = 2t_0\cos(k_y),$$

$$d_2 = (2t_1 \cos(k_y + \phi) + t_0 - \Delta)e^{\frac{ik_x}{2}}$$

$$+ (2t_1 \cos(k_y - \phi) + t_0 + \Delta)e^{-\frac{ik_x}{2}}. \quad (5)$$

The competition between $T$ (controlled by $\phi$) and $P$ (controlled by $\Delta$) symmetries stabilize a topological insulator phase characterized by the Berry-dipole moment $d$, which is defined as the flux of Berry curvature through a half Brillouin zone ($k_x \in [-\pi, \pi]$, $k_y \in [-\pi/2, \pi/2]$). Fig. 3(b) shows the phase diagram in the $(\phi, \Delta)$ plane. A topological region appears when $\Delta < |2t_1 \sin(\phi)|$, within which the dipole moment $d = \pm n$ for $\phi \gtrless 0$ [see details in SM, section 2]. This defines a dipolar Haldane phase, a Berry-dipole counterpart of the familiar Chern insulator.

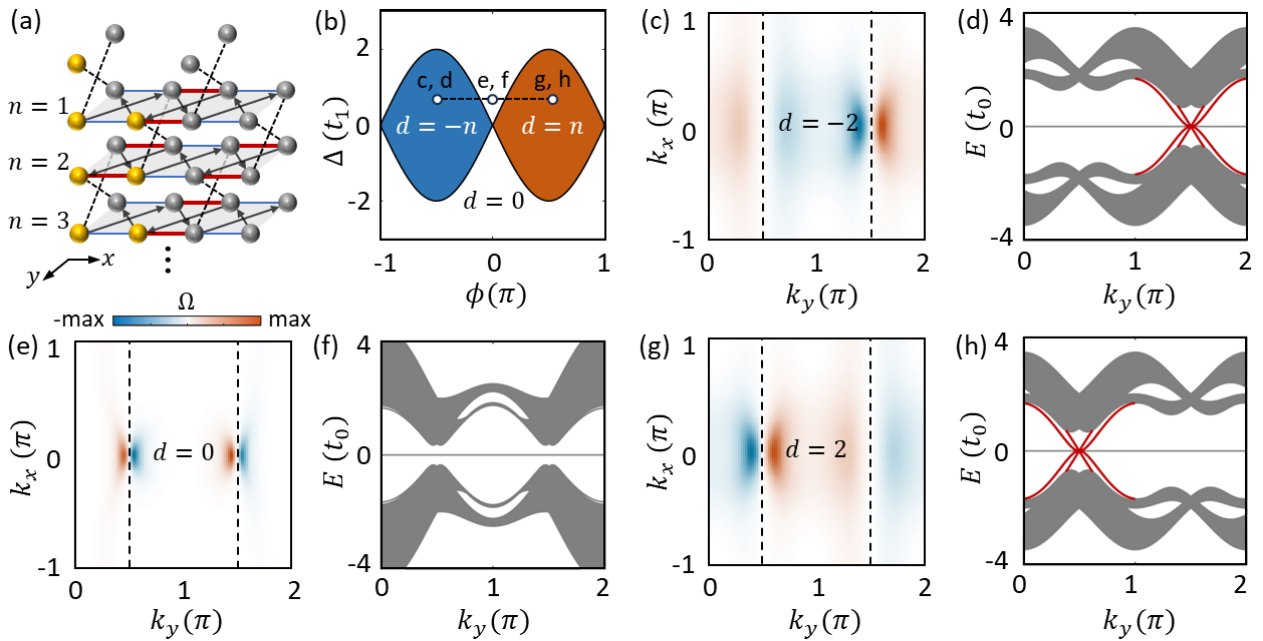

FIG. 3. Dipolar Haldane model and bulk-boundary correspondence. (a) Schematic of the lattice for dipolar Haldane model. The unit cell contains $2n + 1$ sites (yellow spheres). The couplings: diagonal bonds $t_0$ (dashed), complex diagonal bonds $t_1 e^{i\phi}$ (arrowed), and leg bonds $t_0 \pm \Delta$ (red/blue). (b) Phase diagram in the $(\phi, \Delta)$ plane. The shaded region satisfies $\Delta < |2t_0 \sin \phi|$ and hosts a quantized Berry-dipole moment $d = \pm n$. (c-h) The Berry curvature distributions for $d = -2$ (c), $d = 0$ (e), and $d = 2$ (g). The $T$-breaking term $\phi$ is respectively fixed to $-\pi/2$, $0$, and $\pi/2$. The $P$-breaking term $\Delta = 0.6t_1$. The corresponding band spectra for finite lattice in the $x$ direction for $d = -2$ (d), $d = 0$ (f),

and $d = 2$ (h). The parameter $t_1 = 0.3t_0$.

To illustrate the bulk-boundary correspondence, we take $n = 2$ as an example. In a topological phase with $\phi = -\pi/2$ and $\Delta = 0.6t_1$, the Berry curvature [Fig. 3(c)] exhibits a dipole pattern centered at $k_y = \pm\pi/2$ with opposite orientations, giving rise to a quantized dipole moment $d = -2$ upon integration over the half-Brillouin zone. The band spectrum for a finite system along $x$ [Fig. 3(d)] host two pairs of helical edge states near $k_y = \pi/2$, matching the invariant $d = -2$. As we increase $\phi$ from $-\pi/2$ toward $0$ (i.e., weakening the $T$-breaking term), the bulk gap at $k_y = \pi/2$ closes and reopens, driving a transition to a trivial phase at $\phi = 0$ [Figs. 3(e, f)]. In this phase, the Berry curvature shows zero dipole moment, and the edge states vanish. Further increasing $\phi$ to $\pi/2$ reverses the sign of the $T$-breaking term, which closes and reopens the bulk gap at $k_y = -\pi/2$, restoring a topological phase with $d = 2$ [Fig. 3(g)]. Consequently, two pairs of helical edge states emerge, but now localized at the opposite momentum $k_y = -\pi/2$ [Fig. 3(h)]. This full evolution, where the edge-state location flips with the sign of $d$ while their number remains fixed by $|d|$, establishes a bulk-boundary correspondence for the dipole topology. The band structures for other cases with different $n$'s are shown in SM, section 3.

*Oriented helical zero modes.* Furthermore, we couple the dipole system to a pseudomagnetic field to realize helical zero modes, whose existence depends on the orientation of the pseudomagnetic field and number depends on the Berry-dipole moment $n$. We consider a layered Lieb-lattice model shown in Fig. 4(a), where the number of stacked layers encodes the Berry-dipole order $n$. The k-space Hamiltonian satisfies Eq. (1) with

$$d_1 = 2t_0 \cos\left(\frac{k_y}{2}\right),$$

$$d_2 = 2t_0 \cos\left(\frac{k_x}{2}\right) - 2i\Delta \sin\left(\frac{k_x}{2}\right), \quad (6)$$

where $t_0$ is the interlayer coupling (black bond), and red (blue) bonds represent the coupling $t_0 + \Delta$ ($t_0 - \Delta$). A spatial gradient is introduced by setting $\Delta = Bx$ along the $x$ direction, acting as a pseudomagnetic potential $A_z = Bx$ for an effective third dimension ($k_z$). One can have the low-energy Hamiltonian around $(\pi, \pi)$ as

$$d_1 = -t_0 \delta k_y, \quad d_2 = -t_0 \delta k_x - iBx, \quad (7)$$

where $\delta k_{x,y} = k_{x,y} - \pi$. This effective Hamiltonian corresponds to the $k_z = 0$ slice of a 3D Hamiltonian $d_2 = t_0 \delta k_x - i(k_z + A_z)$. The pseudomagnetic field is $\vec{B} = \nabla \times \vec{A} = -B\hat{y}$, while the Berry-dipole moment of the central flat band is $\vec{d} = (0, -n, 0)$.

To illustrate the helical zero modes, we take $n = 2$ as an example. As shown in Fig. 4(b), the linearly decreasing (increasing) $\Delta$ with $x$, represented by blue (red) line, corresponds to the pseudomagnetic field $\vec{B}$ antiparallel (↑↓) [parallel (↑↑)] to the dipole $\vec{d}$. The supercells for these two cases are shown in blue and red blocks in Fig. 4(b). We first plot the band spectrum for $\vec{B} \uparrow\downarrow \vec{d}$ (the supercell in blue block) in Fig. 4(c), where the flat band is isolated without gapless zero modes. In a sharp contrast, for $\vec{B} \uparrow\uparrow \vec{d}$ (the supercell in red block) shown in Fig. 4(d), there appear two pairs of gapless helical zero modes crossing the central flat band and connecting top and bottom bands. We further demonstrate the field distributions of these four helical zero modes in Fig. 4(e), that exhibit Gaussian profiles and localize at the center of the supercell, confirming their nature as bulk topological states.

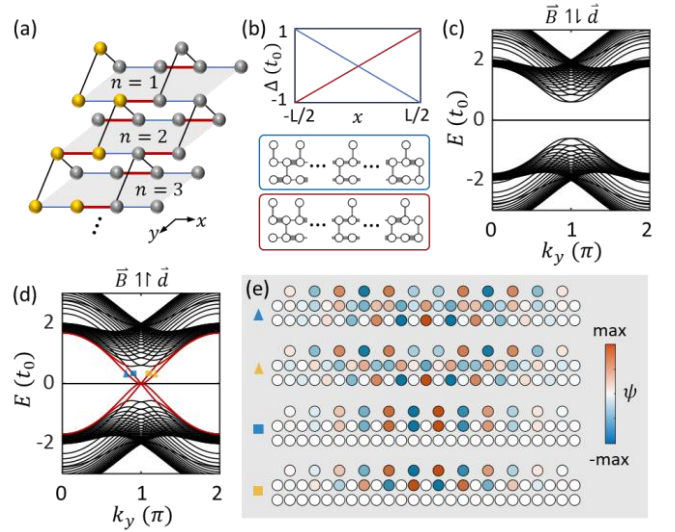

FIG. 4. Oriented bulk helical zero modes in a layered Lieb-lattice model. (a) Schematic of the layered Lieb-lattice model. The number of stacked layers encodes the Berry-dipole order $n$. Black bonds denote $t_0$, while red (blue) bonds denote $t_0 \pm \Delta$, respectively. (b) Spatial profile of the linearly varying coupling imbalance $\Delta = Bx$, inducing a pseudomagnetic field is $\vec{B} = -B\hat{y}$. The supercell configuration for $B \lessgtr 0$ and $n = 2$ is presented in the blue (red) block. The supercell contains $L = 20$ unit cells. (c, d) The band spectra for $\vec{B} \uparrow\downarrow \vec{d}$ and $\vec{B} \uparrow\uparrow \vec{d}$, respectively. Red lines represent helical zero modes. (e) The field distributions for four helical zero modes marked in (d).

*Discussion.* In summary, we have established the quantized Berry dipole as a universal, integer-valued topological invariant for perfectly flat bands in chiral-symmetric (2n+1)-band systems. This higher-order multipole moment governs a suite of topological responses across dimensions, including generalized RTPs, dipolar Haldane insulators with helical edge states, and pseudomagnetic-field-orientation-dependent bulk helical zero modes, as demonstrated through lattice models.

These phenomena are within reach of current experimental platforms. The generalized RTP, featuring bidirectional, "soliton-like" Wannier-center traversal over exactly $n$ unit cells, can be realized in femtosecond-laser-written photonic waveguide arrays, a mature platform for adiabatic topological pumping with precise control over couplings [32–34]. Furthermore, this platform is ideally suited to study the interplay between Kerr nonlinearity and RTP [35,36]. The oriented bulk helical zero modes can be readily realized in tight-binding acoustic or photonic lattices [37–39], providing a pathway to multi-mode, pseudomagnetic-field-selective topological bulk propagation in acoustic/photonic wave systems.

Beyond these signatures, our framework offers a platform for engineering tunable giant quantum metric (always larger than Berry curvature) via increasing the dipole index $n$ in isolated flat bands. Such giant, tunable quantum metric is anticipated to profoundly influence interaction effects, including anomalous coherence lengths in flat-band superconductivity and possibly stabilization of exotic fractionalized or correlated phases driven by geometric contributions [26–29].

†shuzhang@hku.hk